# Relationship between the visibility of political leaders during campaign and the outcome in general elections. A case study for Spain.

J. Estevez ; JJ. Dominguez; M. Graña

*University of the Basque Country. Donostia (Spain)*

## Abstract

In this article, the authors find the evidence that media coverage consisting of 13 online newspapers enhanced the electoral results of right wing party in Spain (Vox) during general elections in November 2019. We consider the political parties and leaders' mentions in these media during the electoral campaign from 1[st] to 10[th] November, and only visibility or prominence dimension is necessary for the evidence.

Keywords: *elections, visibility, media, right-wing*

## INTRODUCTION

In the last decades, we have witnessed a rise of left and right-wing populist parties all over the world. Apart from the socio-economic situation, mass and social media have played a crucial role in all this (Hosch-Dayican, 2016). Recently, a number of scholars have studied the relationship between the media agenda and the electoral success of these parties (Boomgaarden and Vliegenthart, 2007; Walgrave and De Swert, 2004; Kioupkiolis, 2016).

Spanish political parties panorama have suffered a great change in last years. After a long lasted age of what they called two-party system of PSOE and PP (Socialist and Conservative groups, respectively), new political groups have entered with a big presence in the Parliament: Unidas Podemos (left-wing), Ciudadanos (center-wing) and Vox (right-wing).

In last general elections in Spain, on 10[th] November 2019, the far right-wing party Vox obtained a huge increase of votes. In this article, we try to see the effect the of the media on this phenomenon

Populists communication is quite established, which is built around a charismatic party leader. The communication consists on simple, strong and intense language, particularly transmitting decisiveness for radical measures (Bos et al., 2010).

Populists traditional tend and need to call to dramatization and need to generate social tension in order to build up support for the party, and make the population perceive a crisis situation (Albertazzi, 2007).

Normally, common people will never meet any political candidate, so it is through the mass media that voters hear and see the political candidates, and it can therefore be expected that the media has an effect on a voter's perception of candidates (Benoit, Hansen and Verser, 2003; Domke et al., 1997; Esser and Strömbäck, 2014; Mendelsohn, 1996).

We can distinguish three dimensions in media coverage of politicians: prominence, authoritativeness and populism (Bos et al, 2011). The most important dimension is prominence or visibility, which is the amount of media attention for a politician (Watt, Mazza and Snyder, 1993). Prominence makes a small party relevant for voters: when they get more attention, they are possibly perceived as a party that is taken seriously and able to get in power (Bos et al., 2011). It is not necessary the mentions to be positive. Mentions by themselves, translate into popularity (Stray, 2016).

Media attention is often measured by coding newspapers and occasionally television news. Some studies find newspapers to be more influential than television (Walgrave et al., 2004; Lo et al, 2019).

In Spain, there were two general election processes in 2019. First one on 28[th] April and second, on 10[th] November. From one to the other, there was a huge increase in seats by the right-wing party, Vox. In this article, we focus on the prominence of all the political parties in 13 main online newspapers, and try to find an evidence of the electoral ascension of Vox.

The article is divided in following sections: First, in Background, a state of art of the influence of media and political parties (particularly right-wing) is presented. Next, in Methodology, the experiment that was carried out is described. In Results, we detail the outcome of the experiment and finally in Discussion it is argued whether there is any kind of correlation between media coverage and right-wing electoral results.

## BACKGROUND

Scientific bibliography suggests that mass media coverage, as the primary channel through which the electorate receives information about politicians and parties, affects many different aspects of electoral politics (Beck et al., 2002; Norris 2000; Paletz, 1996). Moreover, research works evidence that this effect has a big impact on campaign periods (Aaldering et al., 2018; Ansolabehere et al., 1991; Bos et al., 2011; Gerber et al., 2009).

It appears that public support does not drive media coverage for populist right-wing parties, but that media coverage drives their public support (Boomgaarden and Vliegenthart 2007)

Because proportional representation systems are associated with a greater number of small parties and tend to produce more diverse news (Baum, 2012; Benson, 2009; Kumlin 2001), research confined to such systems is arguably most likely to reflect a model in which media coverage generates support for populist right-wing parties (Murphy, 2018): mass media cover all the political spectrum and tend to oppose extreme and mild opinions. This fact enhances the visibility of populist parties.

Researchers have found some evidence that quantity of media coverage pushes the electoral results of these parties. For instance, (Walgrave and Swert, 2004) find, using time-series data from Belgium, that newspapers and television stations helped to increase the electoral results of the right-wing party by emphasizing political issues owned by them. That study finds evidence suggesting that party and party leader visibility is associated with the electoral outcomes of the parties, but not vice versa. In another study, media coverage was found to be one of the best predictors of electoral success in Denmark's 2007 election (Hopmann et al. 2010; Murphy and Devine, 2018).

A variety of explanations have been suggested for how media cues drive support (Sides and Vavreck, 2013) and some reasons have been clearly identified. First, voters might try to identify which candidates are most likely to win the general election. Media coverage is a form of status conferral and people may interpret media coverage as an indicator of who is electable (Abramowitz, 1989). Second, electors may understand that the candidate with the biggest media coverage is the one with the biggest social support, and that number of public appearances are a reflection of people requests and support (Rothschild and Malhotra, 2014). Finally, there is a large literature within political psychology and communication about the nature of public opinion as a reflection of what is most easily accessible to voters (Zaller, 1992). Beyond the support of candidates, news media is an important force in setting the public agenda. As the press covers a candidate more, voters will be more interested in that candidate (Reuning, 2019).

## METHODOLOGY

The authors developed a software ad-hoc in order to scrape the mentions related to the 5 main political parties in Spain for the general elections on 10th November 2019 (PP, PSOE, Unidas Podemos, Ciudadanos, Vox, which cover from far-left to far-wing ideologies). We scrapped the daily main RSS of the 13 most representative online newspapers in the country. These were: El País, El Mundo, ABC, La Razón, El Español, El Confidencial, ElDiario.es, Diario Público, La Vanguardia, El Periódico de Cataluña, El Correo, El Nacional and Diari Ara. These media are representative of liberal, conservative, leftish and nationalist ideology and generate all together more than 4,000 URL to be analysed during the experiment. The code is housed in Python and is available at GitHub repository[1].

A list of words for each political party was designed, including the name of the group, the leader, and some other related words, which are commonly used by media to refer to specific parties, depending on their tradition. For instance, it is quite usual to refer to the conservatives, PP, as '*Genova*', or to Unidas Podemos as '*los morados*' (*purples*). Moreover, it must be taken into consideration that the governing political party in that moment was PSOE, and so for its list, words like '*the president*', or '*the Government*', were included. This methodology is similar to (Murphy and Devine, 2018), (Walgrave and Swert, 2004) or (Lo et al., 2019)

The software captured the amount of times each word was written on the 13 newspapers, and the URL in which each of one appeared.

The experiment was implemented from 1st to 10th November, which is the campaign period. In that period, three electoral debates where showed on TV, which produced a huge amount of information and mentions the next day.

As for the discussion, we will focus on the difference of general election results on 28th April 2019 and 10th November 2019. In seats, these were the results for each party:

|  | PSOE | PP | Ciudadanos | Unidas Podemos | Vox |
|---|---|---|---|---|---|
| **28th April** | 123 | 66 | 57 | 42 | 24 |
| **10th Nov** | 120 | 88 | 10 | 35 | 52 |

*Table 1: Summary of results in General Elections of 28th April and 10th November 2019, in seats*

And it is also valuable to consider the main poll before elections, on 29th October, carried out by a public institution (Center of Social Research and Studies, CIS) (CIS, 2019).

|  | PSOE | PP | Ciudadanos | Unidas Podemos | Vox |
|---|---|---|---|---|---|
| **CIS 29th Oct** | 133-150 | 74-81 | 27-35 | 37-45 | 14-21 |

*Table 2: Main poll before elections of 10th November (CIS, 2019)*

**RESULTS**

A total of 16,220 mentions were captured. In Table 3 and 4, we show the share of total mentions of parties and political leaders each day. Politicians and parties are related as: Pedro Sánchez (PSOE), Pablo Casado (PP), Albert Ribera (Ciudadanos), Pablo Iglesias (Unidas Podemos), Santiago Abascal (Vox).

|  | PSOE | PP | Ciudadanos | Unidas Podemos | Vox |
|---|---|---|---|---|---|
| **01-nov** | 27,61 % | 27,55 % | 21,52 % | 12,19 % | 11,13 % |
| **02-nov** | 27,33 % | 23,96 % | 21,61 % | 11,82 % | 15,27 % |
| **03-nov** | 25,13 % | 27,68 % | 20,92 % | 10,91 % | 15,35 % |
| **04-nov** | 21,59 % | 24,03 % | 32,31 % | 6,27 % | 15,81 % |
| **05-nov** | 23,06 % | 20,64 % | 27,15 % | 10,80 % | 18,35 % |
| **06-nov** | 18,74 % | 27,78 % | 21,71 % | 12,31 % | 19,46 % |
| **07-nov** | 25,49 % | 24,72 % | 22,18 % | 11,90 % | 15,70 % |
| **08-nov** | 22,24 % | 21,53 % | 25,54 % | 6,70 % | 24,00 % |
| **09-nov** | 21,61 % | 28,71 % | 22,85 % | 8,17 % | 18,66 % |
| **10-nov** | 14,55 % | 18,93 % | 34,14 % | 6,64 % | 25,73 % |

*Table 3: Results of mentions share of political parties*

|         | Sánchez | Casado  | Rivera  | Iglesias | Abascal |
| ------- | ------- | ------- | ------- | -------- | ------- |
| **01-nov** | 40,99 % | 18,57 % | 21,87 % | 13,62 % | 4,95 % |
| **02-nov** | 40,32 % | 12,42 % | 20,81 % | 20,32 % | 6,13 % |
| **03-nov** | 43,59 % | 21,43 % | 13,99 % | 14,87 % | 6,12 % |
| **04-nov** | 42,96 % | 15,00 % | 20,19 % | 11,85 % | 10,00 % |
| **05-nov** | 30,10 % | 15,87 % | 20,66 % | 19,02 % | 14,36 % |
| **06-nov** | 22,02 % | 29,35 % | 15,46 % | 20,84 % | 12,33 % |
| **07-nov** | 32,84 % | 15,44 % | 16,26 % | 24,47 % | 11,00 % |
| **08-nov** | 47,67 % | 10,38 % | 20,75 % | 11,13 % | 10,08 % |
| **09-nov** | 26,74 % | 27,11 % | 19,40 % | 15,05 % | 11,69 % |
| **10-nov** | 23,24 % | 17,09 % | 17,84 % | 11,81 % | 30,03 % |

*Table 4: Results of mentions share of political leaders*

The difference in mentions share from 1$^{st}$ to 10$^{th}$ November was next:

|                | PSOE    | PP      | Ciudadanos | Unidas Podemos | Vox     |
| -------------- | ------- | ------- | ---------- | -------------- | ------- |
| **Difference** | **-47,29%** | **-31,30%** | **58,64%** | **-45,50%** | **131,26%** |

*Table 5: Difference in mentions share of political parties from 1$^{st}$ to 10$^{th}$ November*

|                | Sánchez | Casado  | Rivera  | Iglesias | Abascal  |
| -------------- | ------- | ------- | ------- | -------- | -------- |
| **Difference** | **-43,30%** | **-7,99%** | **-18,43%** | **-13,28%** | **506,44%** |

*Table 6: Difference in mentions share of political leaders from 1$^{st}$ to 10$^{th}$ November*

## DISCUSSION

In this article, we present an experiment in which we measure the prominence or visibility of the political parties, and try to see the influence of media coverage (through online newspapers) on the increase of right-wing party seats. The conclusion of the authors is that it appears to be a clear evidence on the impact of the mentions and electoral results in case of Vox, which needs to be completed in future electoral processes.

In the case of the rest of the leaders and parties, the correlation between electoral results and mentions is not so strong, with the exception of the far left-wing party, Unidas Podemos, which reduced their presence In Spanish Parliament from 42 to 35 seats. Even the correlation is lower than in Vox case, it comes to confirm that extreme ideologies are hugely affected by media visibility and public agenda that they choose.

One of the most important contributions of this article is the research study in case of media influence in Spain, where there is a very scarce bibliography comparing to other European countries. The second contribution is the proof that the chosen online newspaper are a good sample for this kind of studies.

Future work should go towards the inclusion of a sentiment-feeling analysis in this work for next elections, and the incorporation of more polls as an extra variable of information in order to find whether it exists a stronger correlation between visibility and polls.